\documentclass[twocolumn]{aastex631}

\usepackage{xspace}

\newcommand{\Mtwo}{M82 X-2\xspace}
\newcommand{\nustar}{\textit{NuSTAR}\xspace}
\newcommand{\xmm}{\textit{XMM-Newton}\xspace}
\newcommand{\chandra}{\textit{Chandra}\xspace}
\newcommand{\rxte}{\textit{RXTE}\xspace}

\newcommand{\Mone}{M82 X-1\xspace}
\newcommand{\nph}{\ensuremath{N_{\rm ph}}\xspace}
\newcommand{\msun}{\ensuremath{M_{\rm \odot}}\xspace}
\newcommand{\chitwo}{\ensuremath{\chi^2_2}\xspace}
\newcommand{\tseg}{\ensuremath{t_{\rm seg}}\xspace}
\newcommand{\numax}{\ensuremath{\nu_{\rm max}}\xspace}
\newcommand{\asec}{\ensuremath{^{\prime\prime}}\xspace}

\begin{document}

\title{A NuSTAR study of quasi-periodic oscillations from the ultraluminous X-ray sources in M82}

\author[0009-0001-5473-4953]{Hamza El Byad}
\affiliation{INAF-Osservatorio Astronomico di Cagliari}
\affiliation{Department of Mathematics and Computer Science, University of Cagliari, Italy}

\author[0000-0002-4576-9337]{Matteo Bachetti }
\affiliation{INAF-Osservatorio Astronomico di Cagliari}

\author[0000-0001-5420-0719]{Silvia Columbu}
\affiliation{Department of Mathematics and Computer Science, University of Cagliari, Italy}

\author[0000-0001-9054-8712]{Giuseppe Rodriguez}
\affiliation{Department of Mathematics and Computer Science, University of Cagliari, Italy}

\author[0000-0001-7397-8091]{Maura Pilia}
\affiliation{INAF-Osservatorio Astronomico di Cagliari}

\author[0000-0002-8183-2970]{Matthew J. Middleton}
\affiliation{University of Southampton}

\author[0000-0001-5819-3552]{Dominic J Walton}
\affiliation{Centre for Astrophysics Research, University of Hertfordshire}

\author[0000-0002-8147-2602]{Murray Brightman}
\affiliation{California Insitute of Technology}

\author[0000-0001-5857-5622]{Hannah Earnshaw}
\affiliation{California Insitute of Technology}

\author[0000-0001-5800-5531]{Karl Forster}
\affiliation{California Insitute of Technology}

\author[0000-0002-1984-2932]{Brian Grefenstette}
\affiliation{California Insitute of Technology}

\author{Felix F\"urst}
\affiliation{ESA/ESAC}

\author[0000-0002-1082-7496]{Marianne Heida}
\noaffiliation

\author[0000-0001-8688-9784]{Matteo Imbrogno}
\affiliation{INAF-Osservatorio Astronomico di Roma}

\author[0000-0002-6421-2198]{Eleonora Veronica Lai}
\affiliation{INAF-Osservatorio Astronomico di Cagliari}

\author[0000-0003-0976-4755]{Thomas Maccarone}
\affiliation{Department of Physics \& Astronomy, Texas Tech University, Box 41051, Lubbock TX 79409-1051 USA}

\begin{abstract}

The study of quasi-periodic oscillations in X-ray binaries provides valuable insights into the physics of accretion around compact objects. The M82 galaxy hosts two ultraluminous X-ray sources (ULXs), one of which is suspected to harbor an intermediate-mass black hole. Using 39 
\nustar observations acquired between 2014–2024, we investigate the aperiodic X-ray variability in M82. In particular, we study in detail the evolution of the QPO from M82 X-1 in the range 20--300 mHz. We do not find additional timing features in the data, besides a frequent broad noise component at lower frequencies. The QPO behaves similarly to other classes of low-frequency oscillations in accreting compact objects, both black holes and neutron stars.
\end{abstract}


\section{Introduction} \label{sec:intro}

One of the most intriguing phenomena in the accretion processes of X-ray binaries is the presence of quasi-periodic oscillations (QPOs; see \citealt{ingram2019review, van1989fourier} for reviews). They are nearly periodic variations manifesting in the form of relatively coherent oscillations in the X-ray emissions from accretion disks surrounding neutron stars (NS) and black holes (BH).
These oscillations have been observed across a wide range of frequencies in X-ray binary systems. They are generally categorized into two types: low-frequency QPOs (LFQPOs), which have been detected in the majority of black hole binaries (BHBs) observed with Rossi X-ray Timing Explorer (\rxte; \citealt{bradtXrayTimingExplorer1993}), with a frequency that can extend up to $\sim 30 \mathrm{~Hz}$  \citep{Belloni2005}, and up to $\sim 60 \mathrm{~Hz}$ in neutron-star binaries (NSBs;\citealt{vanderklisRapidXrayVariability2006}); and high-frequency QPOs (HFQPOs or kHz QPOs), observed between 100 and 500 $\mathrm{Hz}$ in BHBs and from several hundred $\mathrm{Hz}$ to over $1 \mathrm{kHz}$ in NSBs with weak magnetic fields \citep{motta2017links}. These variations highlight some of the differences in the variability of BHBs and NSBs, reflected in the distinct variability patterns of these systems.
By studying these oscillations, researchers can gain insight into the fundamental physics of compact objects, including their masses, radii, and spin  \citep{klu2006}, and the differentiation between different types of compact objects, such as BHs and NSs. For example, certain frequency patterns may indicate specific properties unique to NSs or quark stars \citep{Shaposhnikov2012}. The characteristics of QPOs can also offer critical information about the dynamics of matter accretion, the structure of the inner accretion disk, and the strong gravitational environments near compact objects \citep{remillard2006xray, belloni2012hfqpo, zhang2006hfqpo}. High-frequency QPOs, in particular, have been suggested to originate from relativistic precession \citep{ingram2011model} or from non-linear resonances between orbital epicyclic modes—specifically the radial and vertical oscillation frequencies—within the inner accretion disk \citep{kluzniak2001strong, abramowicz2001precise, motta2016review}. Thus, these oscillations serve as a valuable tool for probing the physics of accretion and the fundamental properties of compact objects.

The M82 galaxy harbors a number of ultraluminous X-ray sources (ULXs), off-nuclear X-ray sources where the accretion process implied by their luminosity exceeds the theoretical Eddington limit \citep{Kaaret2017, King2023NewAR}.
Two in particular have received considerable attention in the last years: \Mtwo (the first ULX found to be powered by a NS) and \Mone (a candidate intermediate-mass black hole, IMBH).
This cigar-shaped galaxy has a star formation rate of approximately 10 M$_{\odot}$ yr$^{-1}$ \citep{Iwasawa2021} and lies at a distance of 3.6 Mpc \citep{Freedman1994}, and has the distinguishing feature of having a low-luminosity active galactic nucleus (LLAGN) as reported by \citet{Matsumoto2001}.

\Mone has the ability to achieve X-ray luminosities up to $\sim10^{41}$ erg s$^{-1}$, and it is typically the brightest X-ray source in M82. However, observing this unique source presents significant observational challenges due to  its complex environment. It resides nearby several bright X-ray sources, most notably \Mtwo, X-3 and X-4 \citep{brightmanSpectralEvolutionUltraluminous2020}.

In particular \citet{Bachetti2014b} identified \Mtwo, the second source in order of luminosity, as the first ultraluminous X-ray pulsar (PULX or ULXP)  ever discovered, with a spin period of 1.37\,s and a 2.5-day orbit with a semi-major axis of 22-light seconds. A long-term tracking of the pulsar revealed that the orbit is decaying at a rate of $-5.69(24)\cdot10^{-8}$\,d/d \citep{Bachetti2022}.
\Mtwo resides approximately 5\asec away from \Mone, which causes substantial contamination issues in both spectral and timing analyses. Currently, only the \chandra X-ray Observatory \citep{weisskopfOverviewPerformanceScientific2002} has the angular resolution (~0.5\asec) sufficient to spatially resolve these sources. However, \chandra has a limited effective area and energy range (0.3-10 keV), which restricts detailed spectral studies of the hard X-ray emission crucial for understanding accretion flows.

Timing analysis of M82's X-ray emission has provided major evidence that \Mone is a promising IMBH candidate. Initial evidence emerged from the detection of a QPO at 54 mHz associated with \Mone using \xmm data by \citet{Strohmayer2003}.
Building on this research, \citet{Mucciarelli2006} and \citet{Dewangan2006} reported QPOs spanning between 50–166 mHz using combined \xmm and 
\rxte archival observations.
By examining the relationship between this QPO frequency and the photon index of the energy spectrum, they estimated the BH mass of \Mone to be between 25 and 520 solar masses. The most compelling evidence for \Mone harboring an IMBH came from \citet{Pasham2014}, who reported stable twin-peak QPOs with a 3:2 frequency ratio at 3.3 and 5.1 Hz using 
\rxte data.
This frequency pattern, which is characteristic of high-frequency QPOs in stellar-mass BH binaries but scaled down to lower frequencies, is potentially a signature of a BH of approximately 400 solar masses. This would place \Mone firmly in the intermediate-mass range -- a critical  ``missing link'' between stellar-mass and supermassive BHs. \citet{atapinUltraluminousXraySources2019} analyzed the X-ray power density spectra of several ULXs, including \Mone, and found that the QPO frequencies are anti-correlated with the level of flat-topped noise, suggesting that mass accretion rate variations influence the observed variability (see also \citealt{Middleton2015a}).

Adding to the complexity in the high-energy regime in M82, \chandra observations revealed low-frequency QPOs (3-4 mHz) from \Mtwo \citep{feng2010discovery}. These findings underscore the diversity of ULXs within M82, each with different timing properties, thereby offering additional avenues to study varied accretion mechanisms.


In this work, we analyze observational data acquired with NASA's \nustar \citep{harrisonNuclearSpectroscopicTelescope2013a} satellite spanning from 2014 to 2024, to investigate the variability of the X-ray flux in the M82 galaxy.

\section{Data reduction } \label{sec:datared}

\subsection{\nustar}
We used data from all available \nustar observations of the M82 galaxy between $2014$ and $2024$, reduced with the standard pipeline (\texttt{nupipeline} from NuSTARDAS\footnote{\url{https://heasarc.gsfc.nasa.gov/docs/nustar/analysis/nustar_swguide.pdf}}) from the High Energy Astrophysics Science Archive Research Center (HEASARC), and selected photons from a region of 70\asec around the position of the source.
The full list of observation is in Table~\ref{tab:rnqpo}.
Our data consist of individual events that are timestamps of when a photon reached the detector and their associated properties, such as the photon energy.
Each timestamp was corrected from local time to the barycenter of the Solar System using the \texttt{barycorr} FTOOL, using the ICRS coordinates of \Mtwo 09:55:51.040 +69:40:45.49\footnote{Data were barycentered in order to allow the study of aperiodic variability from \Mone and pulsations from \Mtwo. A 5\asec mismatch is irrelevant for the study of slow variability, but it might be detectable in precise pulsar timing.} \citep{kaaret62DayXRay2006} and using the DE430 JPL ephemeris\footnote{\url{https://naif.jpl.nasa.gov/pub/naif/generic_kernels/spk/planets/aareadme_de430-de431.txt}}.

We first computed the raw power-spectral density (PSD) for each observation check for telemetry issues and to define an adequate frequency grid for subsequent fitting. 
The primary analysis used only events recorded inside  Good-Time Intervals (GTIs) with full star tracker visibility (mode-01 data). For a limited part of the study, we made use of photons from intervals with limited star tracker coverage (the SCIENCE\_SC mode, or mode-06, data). 
We used the tool \texttt{nusplitsc} to split the mode-06 data into intervals with single star tracker combinations; we moved the extraction region for events in each sub-interval to adapt to the position of the PSF centroid; we extracted the events from the source region; and finally, we merged the events together with the mode-01 data.
Given the circumpolar position of the source and \nustar's near-equatorial orbit, this procedure nearly doubled the number of photons available for sensitive QPO searches, at the cost of increasing the red noise level. Consequently, we used the merged dataset only to increase the sensitivity of our search for new features at high frequencies where the red noise was negligible.

We ran the analysis using Stingray \citep{huppenkothenStingrayModernPython2019}, a Python library built to perform time series analysis, providing implementations of the most advanced spectral timing techniques available in the literature.

Raw data do not provide the photon energy directly. However, in \nustar the energy channel number (PI) and the central energy $E_p$ of the channel are related by the simple formula $E_p \mathrm{(keV)}=1.62 + 0.04\,\mathrm{PI}$.
This is accounted for automatically when loading data in Stingray.

\begin{figure}[tb]
\includegraphics[width=\linewidth]{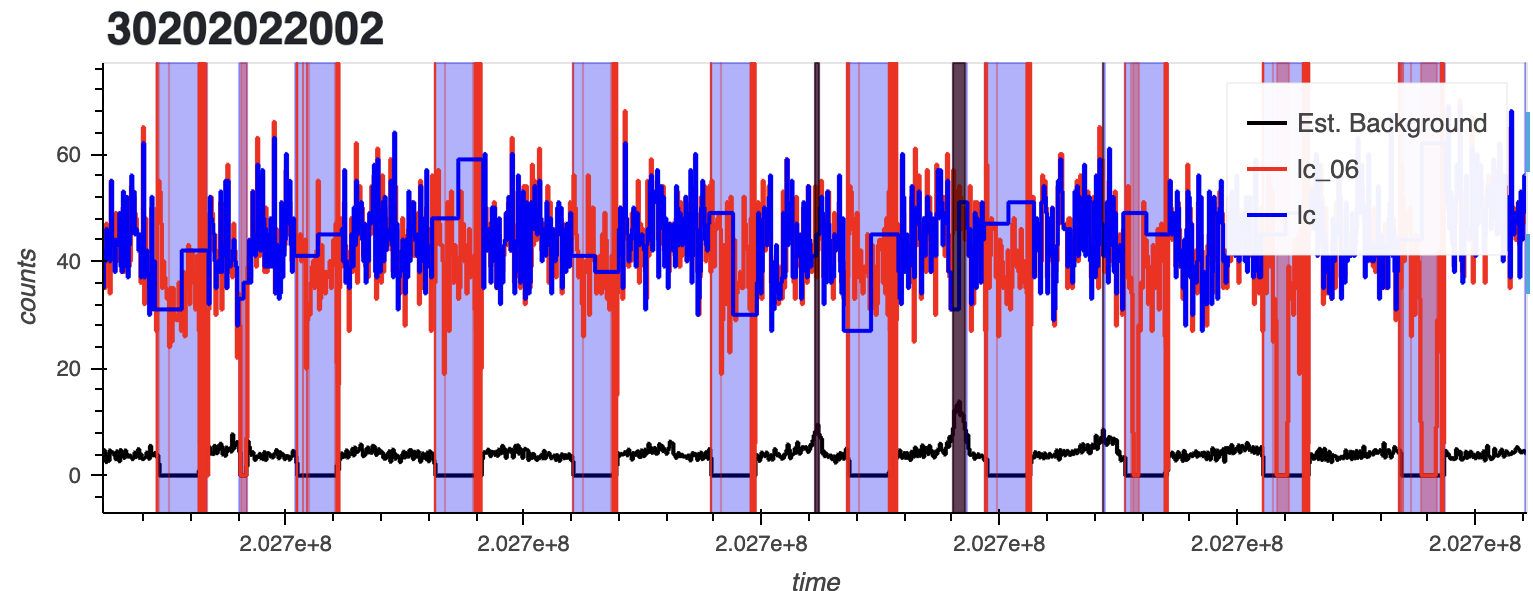}

\caption{Example of the GTI cleaning process. We plot a rescaled background light curve and the source light curve, and eliminate intervals where the background light curve is above 10 \% of the mean level. Vertical black bands indicate these bad intervals, while blue bands indicate the standard bad time intervals due to occultation or poor star tracker coverage.
Red data points are taken in intervals with poor star tracker coverage. We will not consider them when modeling the PDS, but we will use them to improve the detectability of unseen features.\protect}
\label{fig:lcclean}
\end{figure}

We plotted light curves of the source region of all observations, and background light curves containing all photons further than 100\asec from the source.
This criterion is different from the one usually employed for spectral analysis, where a background region is chosen in a large circular region of the FOV devoid of sources. In our case, these background light curves were meant to catch flares in the background, corresponding to anything from particles hitting the telescope or other increased environmental background (e.g. approaching the SAA), and we needed to gather all the photons we could at a reasonable distance from the source region.
We rescaled the background light curve by multiplying by the ratio of pixels inside the source and background regions.
Whenever the rescaled background light curve had flares reaching more than $\sim10$\% of the source mean flux, we rejected that time interval and excluded it from the good time intervals (GTI). See Fig.~\ref{fig:lcclean} for an example.

\subsection{\chandra}
Used a script based on \texttt{astroquery} \citep{ginsburgAstroqueryAstronomicalWebquerying2019} to select all \chandra observations of M82 executed within 1 day of a \nustar ObsID in the same field. 
The search returned the coincidences in Table~\ref{tab:xmatch}. 
We downloaded the \chandra data of these observations and ran the \texttt{chandra\_repro} script distributed with the CIAO software \citep{fruscioneCIAOChandrasData2006} to produce level-2 cleaned event files.
We barycentered the event files similarly to what we did for \nustar data. 
We selected photons from circular or oval regions approximating the shape of \Mone and \Mtwo in the images, typically 1-3 arcseconds wide. The change of shape was particularly accentuated in off-axis observations, that in turn were less affected by pile-up. \Mone and \Mtwo were never overlapped, nonetheless, allowing an easy separation of the photons from the two sources. 
\Mtwo had a slight overlap with the very nearby M82 X-3 and X-4, which are however usually fainter than \Mtwo if the latter is not in a low state \citep{brightman60DaySuperorbital2019}.

\section{Timing analysis } \label{sec:timing}
\subsection{Statistical properties of the periodogram}

Periodogram-based methods are commonly used as a nonparametric approximation of the power density spectrum (PDS). Given the Discrete Fourier Transform (DFT) components $a_i$ of a light curve $x_n$ ($n=0, \dots, N$)

\begin{equation}
	a_{i}= \frac{1}{N}\sum_{n=0}^{N-1} x_{n} e^{-j 2 \pi n i/N }, \quad (i= -N/2, \ldots, N/2-1 )
	\label{an}
\end{equation}
the periodogram is defined as $P_i=| a_i |^2$.

In X-ray Astronomy, it is common to use the normalization from \citet{leahy1983searches}, where the periodogram defined above is multiplied by a factor $2/\nph$, where \nph is the number of photons in the light curve. With this normalization, the powers of a periodogram of pure white noise follow a $\chitwo$ distribution, allowing for an easy identification of outliers.
A common procedure to limit the noise of the periodogram is based on power averaging, either of $W$ nearby bins from the same periodogram, or $M$ periodograms from different segments of the data (the so-called Bartlett periodogram, from \citealt{bartlett1950periodogram}), or using both methods. It is easy to demonstrate that the effect of averaging $MW$ noise powers leads to a normalized $\chi^2_{2MW}/MW$ distribution, which resembles more and more a Gaussian distribution with width $\sigma=2/\sqrt{MW}$ \citep{van1989fourier} as the number of averaged powers increases.

\begin{figure}
    \centering
    \includegraphics[]{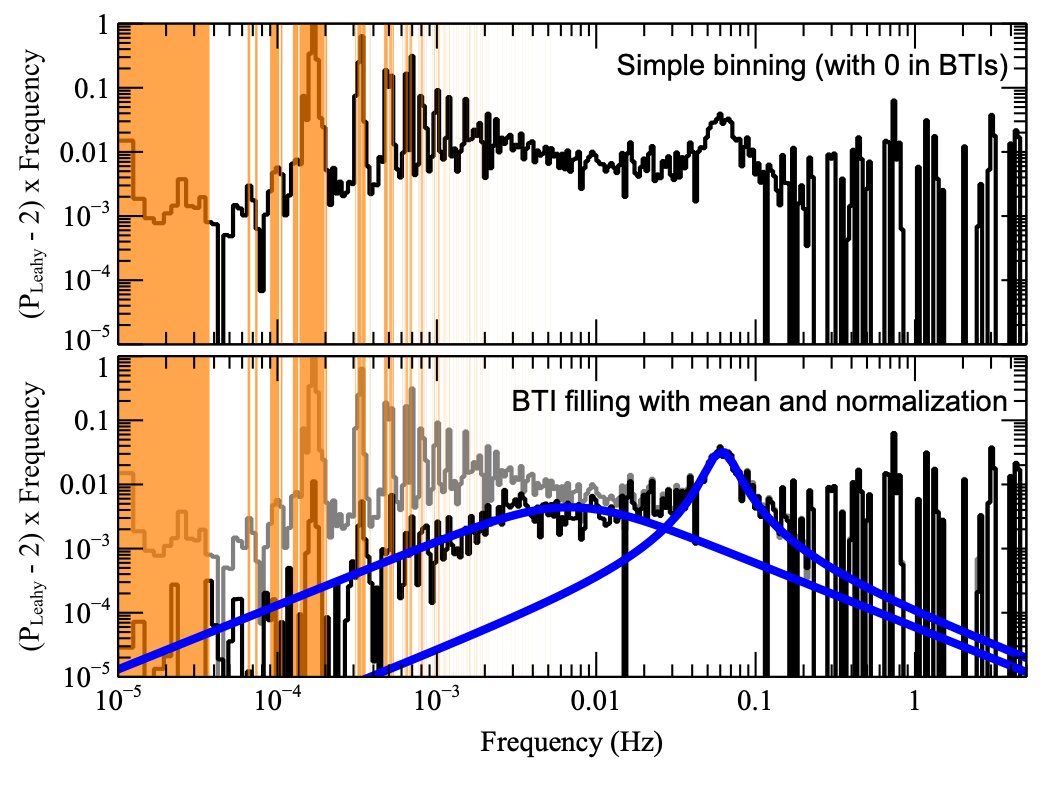}
    \caption{Example analysis using obsid 80002092006. Data are cleaned as described in Section~\ref{sec:periodogram}. Orange bands show the notch-filtered frequencies. The top panel shows the periodogram of the light curve when BTIs contain zeroes, and the lower panel shows the result when BTIs are filled with the mean counts per bin. It is clear that notch filtering is still needed in this non-ideal case. The two blue curves are the two best-fit Lorentzians for the red noise component and the QPO. }
    \label{fig:806}
\end{figure}

The Bartlett periodogram has the additional major advantage of being applicable to observations containing missing data, for example due to Earth occultation, South-Atlantic Anomaly passes, high background, and so on.
In X-ray observations, these ``bad intervals'' are usually eliminated from the observations during the data reduction procedure, and good observing conditions are encoded in a GTI list in the same FITS files of the data.
The Bartlett periodogram can then be chosen so that one or more intervals of duration $\tseg$ fit inside the typical length of GTIs.

The powers $I_j$ of a periodogram containing signal are distributed following a $\chi_{2 MW}^{2}$ around the real spectrum $S_j$ \citep{barret2012maximum}.
\begin{equation}
	    I(f_{j})= \frac{S(f_{j})}{2MW}X
	\label{fen2m}
\end{equation}
where $X$ follows a $\chi_{2 MW}^{2}$ distribution.

From the properties of the $\chi_{n}^{2}$ distribution, the log-likelihood formula in this general case can be derived as follows:
\begin{equation}\label{eq:loglike}
	\log\mathcal{L} = \frac{-\nu}{2} \sum_{j=1}^{N-1}\left\{\frac{I_{j}}{S_{j}}+\ln S_{j}+\left(\frac{2}{\nu}-1\right) \ln I_{j}+c(\nu)\right\}
\end{equation}
where $\nu = 2MW$ is the number of degrees of freedom and  $c(\nu)$ is a constant for fixed $\nu$.
Since minimization algorithms are more common than maximization ones, maximum-likelihood fitting procedures usually consist of minimizing the quantity $-2\log\mathcal{L}$.

The Bartlett periodogram calculated from a typical Fast Fourier Transform \citep{cooleytukey1969} has a major limitation in terms of frequency resolution, limited to $\Delta\nu=1/	\tseg$. This also means that we cannot investigate frequencies lower than $\Delta \nu$.
Some techniques to improve the frequency resolution, such as interbinning or Fourier interpolation \citep{ransomFourierTechniquesVery2002}, can be used, but at the cost of altering the statistical properties of the data and making the fitting and interpretation of results less robust.

On the other hand, a single periodogram of the whole time series has a much better frequency resolution and sensitivity to low frequencies (as now the resolution is $\Delta\nu=1/t_{\rm obs}$), but it is affected by the missing data and contains a large number of (typically) low-frequency peaks that correspond to the variability introduced by the visibility windows.
If the bad time intervals (BTIs) are very small (e.g. less than 1\% of the data), it is customary to add some white noise to fill-up the intervals and have a final periodogram with all the desired statistical properties, at the expense of some minor loss of sensitivity.
However, in our observations, the bad intervals are comparable in length with the good ones, which would imply simulating about half the data, which is unacceptable for our purposes.

For this work, we devised a treatment for the periodogram that limits the effects of windowing while maintaining most of the statistical properties of the periodogram (see Appendix \ref{sec:periodogram} for details).
An example of one of our cleaned periodograms is depicted in Figure~\ref{fig:806}.

\subsection{Model construction}

The periodograms of X-ray binaries show a variety of variable phenomena, and can be conveniently modeled through a composition of Lorentzian components \citep{belloniUnifiedDescriptionTiming2002}. For these functions, we use the definition:

\begin{equation}\label{eq:lorentz}
P(\nu) = \frac{A_{0} \left(\frac{w}{2}\right)^2}{\left(\nu - \nu_{0}\right)^{2} + \left(\frac{w}{2}\right)^{2}}
\end{equation}
where $\nu_{0}$ is the centroid frequency or the frequency at the peak of the signal, $w$ represents the full width at half maximum (FWHM), and $A_{0}$ is the amplitude of the signal.

The use of Lorentzian components is mostly phenomenological given the symmetry properties of these functions, even though it is originally rooted in the fact that Lorentzians are the Fourier transforms of exponentially decaying oscillations, a common phenomenon in nature.

These Lorentzian components can be characterized through three main quantities.
First of all, the \textbf{characteristic frequency} (\numax):
$$
\numax = \sqrt{\nu_0^2 + \left(\frac{w}{2}\right)^2},
$$
that represents the peak of the Lorentzian component in a $\nu P$ vs $\nu$ plot and the frequency at which the Lorentzian contains the most power per logarithmic frequency interval \citep{belloniUnifiedDescriptionTiming2002}.
This quantity is very close to $\nu_0$ for coherent QPOs, while it departs considerably from it for broadband red-noise components.
It is particularly useful when considering the evolution of timing features, as it is common to observe a broadband noise component evolve into a QPO. The characteristic frequency, in this case, makes a smooth transition that would not be as clear when using the central frequency \citep[e.g.][]{motta2016review}.
The second important quantity is the \textbf{Quality factor (\(Q\))}: Defined as \(Q = \frac{\nu_{0}}{\mathrm{HWHM}} = \frac{2\nu_{0}}{w}\), it measures the signal's coherence\footnote{Many works in the literature divide by the full width at half maximum, so their values of $Q$ would be 50\% of the ones calculated here}. Commonly, Lorentzian components with \(Q > Q_{lim}\) are classified as QPOs, while those with \(Q < Q_{lim}\) are considered broadband peaked noise, with different choices of $Q_{lim}$ in different papers. In this work we will use $Q_{lim}=2$ as the boundary between a QPO and a broadband component.
However, it will become clear that the feature we identify as QPO can sometimes have a low coherence.
Finally, we can define
the \textbf{(fractional or absolute) root-mean-squared (rms) amplitude}: A measure of the signal's strength, which depends on the source flux. It is proportional to the square root of the integrated power contributed by the QPO to the periodogram. Using eq.~
\ref{eq:lorentz}, the rms amplitude can be calculated as the square root of the integral of $P(\nu)$ normalized in the desired rms units (fractional, e.g.  \citealt{belloniAtlasAperiodicVariability1990} or absolute, in counts per second). Since power is only calculated at positive frequencies, assuming $P(\nu)$ was fit in Leahy normalization, the rms amplitude can be calculated as:
\begin{equation}
    \mathrm{rms} = \sqrt{F \int_{0}^{\infty} P(\nu) d\nu} =
    \sqrt{A_0F
    \left(
        \pi/2 - \tan^{-1} \frac{-\nu_0}{w/2}
    \right)
    }
\end{equation}
where $F$ is the conversion factor between the Leahy normalization and the desired rms units.

\begin{figure}[htbp]
\includegraphics[width=\linewidth]{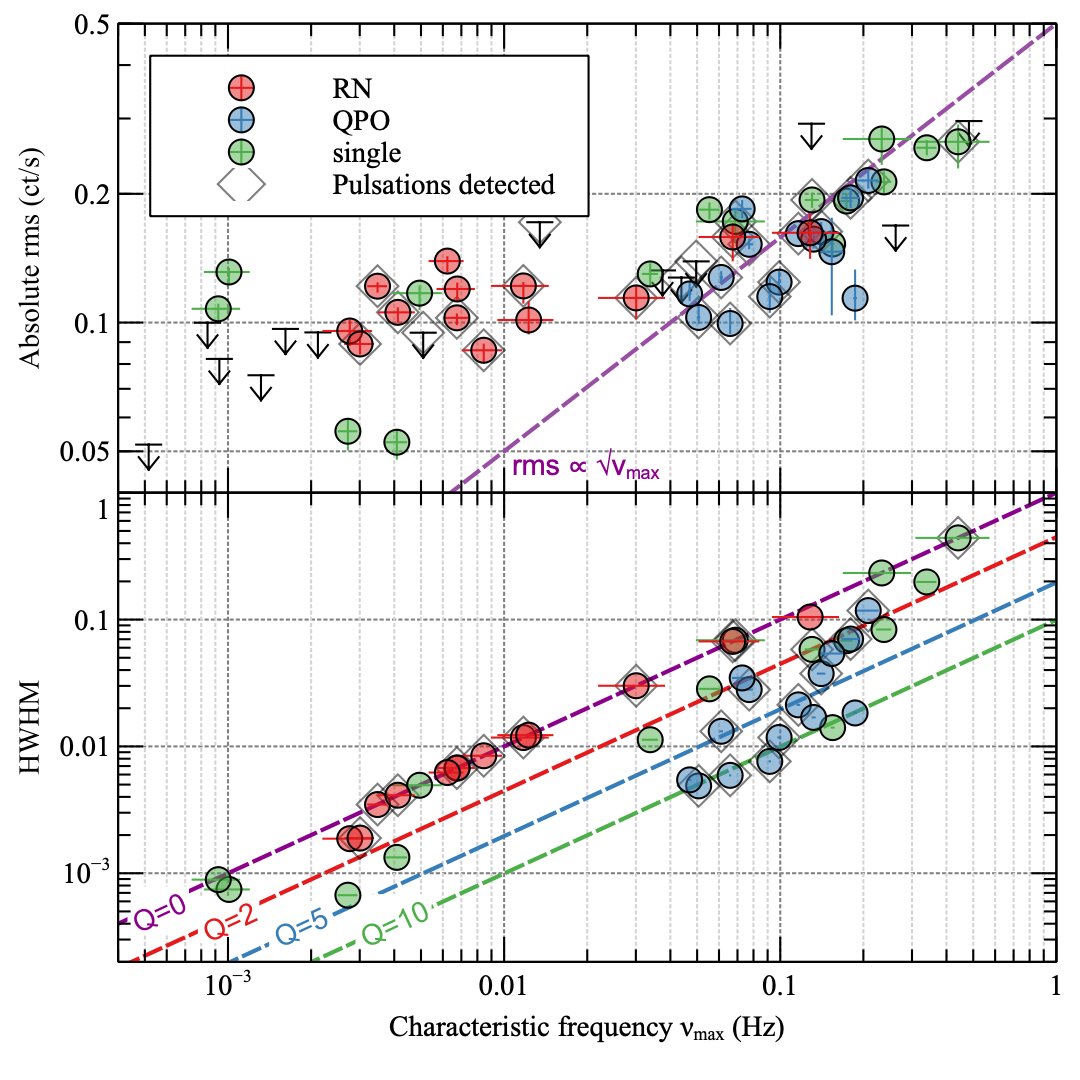}
\caption{(Top) rms versus frequency for the power spectral features fit in Section~\ref{sec:intro} and (Bottom) HWHM versus frequency for the same features. There does not seem to be a systematic pattern in the appearance of these timing features and pulsations.
We calculated 1-$\sigma$ uncertainties and 3-$\sigma$, upper limits through the bootstrap procedure in Section~\ref{sec:bootstrap}.
The identification of the two features is often difficult if only one of them is present in the data, but a simple criteria to distinguish them seems to emerge from this visualization, with the QPO, regardless of its Q factor in a given observation, following a $\mathrm{rms}\propto\nu^{1/2}$ law (dashed line) and generally having a frequency above 0.02 Hz.}
\label{fig:rms}
\end{figure}

\subsection{Inference}\label{sec:bootstrap}
Our periodogram modeling consists of two main steps:
a maximum likelihood estimation (MLE) of the best fitting model, following \citet{barret2012maximum}, and a parametric bootstrap technique to evaluate uncertainties.
The MLE is done with the \texttt{stingray.modeling} package \citep{huppenkothenStingrayModernPython2019}.
The procedure consists of minimizing the negative loglikelihood in Equation~\ref{eq:loglike}, using a model power spectrum.
We employ an optimization algorithm that supports bounds, like the limited-memory Broyden–Fletcher–Goldfarb–Shanno scheme (L-BFGS; \citealt{byrdLimitedMemoryAlgorithm1995}) to maximize the likelihood function, thereby obtaining parameter estimates.
We fit the periodograms with one or two Lorentzians.
When in doubt about the most appropriate number of components, we use the Akaike Information Criterion \citep{akaikeNewLookStatistical1974} to determine whether a model is best described by one or two Lorentzians (with threshold $\Delta$AIC=2).
For the Lorentzian components, we use an Astropy model (\texttt{Lorentz1D}) with the same parameters as Equation~\ref{eq:lorentz}, and we fit an additional constant with starting value 2, the expected white noise level in the Leahy normalization.
For the initial values of the Lorentzians, we use an interactive interface to build a reasonable starting model, but then we do not set boundaries to the parameters other than being positive definite, and the FWHM being more than 0.001 Hz.

Once we obtain a best-fit model, we use a parametric bootstrap to evaluate the uncertainties, which involves the following steps:
\begin{enumerate}
     \item\label{it:random}  \textbf{Generate random powers}: Randomly simulate powers from the best-fit model, ensuring that they follow a distribution of $\chi ^{2}_{2M}/2M$, where $M$ is the number of averaged powers in each bin of the original periodogram. These simulated powers should scatter around the best-fit model.
    \item\label{it:fit} \textbf{Fit a model to the random powers}, starting from random parameters distributed within 10\% of the real parameters.
    \item  \textbf{Bootstrap} procedure:
        Repeat steps \ref{it:random}--\ref{it:fit} 1,000 times, recording the fit parameters in every iteration (a bootstrap distribution is obtained for each parameter)
    \item  \textbf{Parameter estimates}: use the appropriate percentiles from the bootstrap results of the parameters to compute relevant statistics such as means, two-sided standard errors, and confidence intervals around estimates. These provide information on the uncertainty associated with the parameter estimates based on the simulated powers. To take into account cases where the fit swaps the two Lorentzian components, we always order them by central frequency $\nu_{0}$ and, if they both go to zero, by characteristic frequency \numax.
    We consider an outlier any point at more than 5 median absolute deviations (MADs) from the median.
\end{enumerate}

\begin{deluxetable*}{cccccccccc}
\tablewidth{0pt}
\tablecaption{Summary table with all the \nustar datasets in this work, containing best-fit values and $1-\sigma$ uncertainties for the red noise and QPO components obtained in Section~\ref{sec:bootstrap} and information on the observing conditions. \label{tab:rnqpo}}
\tablehead{
\colhead{ObsID} & \colhead{Date} & \colhead{Exposure} & \colhead{rms$_{\rm rn}$}
& \colhead{$\nu_{\rm max, rn}$}
& \colhead{rms$_{\rm qpo}$}
& \colhead{$\nu_{\rm 0,qpo}$}
& \colhead{$w_{\rm qpo}$}
& \colhead{$Q_{\rm qpo}$}
& \colhead{Notes}\\
\colhead{} & \colhead{MJD} & \colhead{ksec} & \colhead{ct s$^{-1}$}
& \colhead{mHz}
& \colhead{ ct s$^{-1}$}
& \colhead{mHz}
& \colhead{mHz}
& \colhead{}
& \colhead{}
}
\startdata
30101045002 & 57493.3 & 195.5 & 0.122(4) & 3.48(29) & 0.161(6) & 114.3(14) & 42(4) & 5.4(6) & Q, P \\
30202022002 & 57542.9 & 40.3 &  &  &  &  &  &  & BF \\
30202022004 & 57570.7 & 48.2 & 0.069(30) & 2.1(12) &  &  &  &  &  \\
30202022008 & 57599.0 & 43.7 & 0.055(24) & 0.9(7) &  &  &  &  & BF, SF \\
30202022010 & 57619.4 & 44.1 & 0.055(20) & 1.3(6) &  &  &  &  & BF \\
30502020002 & 58691.9 & 93.0 &  &  & 0.256(11) & 275(16) & 400(50) & 1.39(27) & Q \\
30502020004 & 58701.9 & 91.5 & 0.135(9) & 0.98(20) &  &  &  &  & SF \\
30502021002 & 58918.1 & 86.3 &  &  & 0.194(8) & 116(5) & 116(15) & 2.01(34) & Q, P \\
30502021004 & 58929.2 & 79.8 & 0.162(21) & 128(35) & 0.114(19) & 185.7(25) & 36(10) & 10.1(29) & BF, Q \\
30502022002 & 59000.8 & 90.9 & 0.120(8) & 6.8(11) & 0.117(8) & 46.6(6) & 10.9(19) & 8.6(16) & SF, Q \\
30502022004 & 59012.6 & 99.2 & 0.096(7) & 2.8(6) & 0.184(9) & 64(4) & 69(11) & 1.8(4) & BF, Q \\
30602027002 & 59311.9 & 73.6 & 0.074(11) & 5.1(23) & 0.196(12) & 165(8) & 140(27) & 2.4(6) & Q, P \\
30602027004 & 59325.3 & 71.6 & 0.159(19) & 67(17) & 0.115(12) & 91.3(8) & 15.2(33) & 11.0(27) & BF, Q, P \\
30602028002 & 59215.2 & 69.0 &  &  & 0.192(8) & 159(6) & 136(18) & 2.3(4) & BF, Q \\
30602028004 & 59226.9 & 70.8 & 0.038(14) & 0.52(25) &  &  &  &  & BF \\
30702012002 & 59504.0 & 128.3 & 0.106(6) & 4.1(6) & 0.163(9) & 135(4) & 75(11) & 3.6(6) & Q, P \\
30702012004 & 59674.3 & 124.4 & 0.13(5) & 130(50) & 0.15(4) & 143(9) & 110(40) & 2.7(12) & Q \\
30901038002 & 60110.6 & 128.1 & 0.108(7) & 0.92(18) &  &  &  &  & BF, SF \\
31001019002 & 60657.4 & 133.9 & 0.089(5) & 3.0(4) & 0.215(11) & 171(10) & 240(40) & 1.46(31) & BF, Q, P \\
50002019002 & 57037.9 & 32.8 &  &  & 0.130(8) & 31.9(16) & 22(4) & 2.8(7) & BF \\
50002019004 & 57041.8 & 168.2 & 0.101(11) & 12.3(28) & 0.114(9) & 36(14) & 14.1(20) & 5.2(27) & BF, SF, Q \\
80002092002 & 56680.5 & 67.4 & 0.130(18) & 12(4) & 0.095(19) & 43(16) & 6.7(15) & 12(8) & Q, P \\
80002092004 & 56682.8 & 92.3 & 0.122(11) & 11.7(28) & 0.103(11) & 50.4(6) & 9.8(17) & 10.3(20) & BF, Q, P \\
80002092006 & 56685.5 & 321.1 & 0.103(4) & 6.7(7) & 0.127(4) & 59.7(7) & 26.3(22) & 4.5(4) & BF, Q, P \\
80002092007 & 56692.2 & 319.3 & 0.086(6) & 8.4(14) & 0.152(5) & 71.9(16) & 56(5) & 2.56(29) & BF, Q, P \\
80002092008 & 56698.8 & 35.2 & 0.172(19) & 68(19) &  &  &  &  & BF, Q, P \\
80002092009 & 56699.5 & 119.6 & 0.114(12) & 30(8) & 0.124(8) & 98.3(12) & 23(4) & 8.4(15) & BF, Q, P \\
80002092011 & 56719.7 & 114.5 & 0.118(15) & 49(17) & 0.099(9) & 65.6(5) & 11.8(17) & 11.1(17) & BF, SF, Q, P \\
80202020002 & 57413.8 & 37.9 & 0.079(18) & 1.6(6) &  &  &  &  & SF \\
80202020004 & 57441.7 & 32.4 & 0.269(35) & 230(60) &  &  &  &  & BF \\
80202020006 & 57483.4 & 31.8 &  &  & 0.184(9) & 47.6(30) & 56(9) & 1.7(4) & BF, SF \\
80202020008 & 57502.8 & 41.7 &  &  & 0.153(12) & 154.1(22) & 28(6) & 10.0(26) & BF \\
80202020008 & 57502.8 & 41.7 &  &  & 0.153(12) & 154.1(22) & 28(6) & 10.0(26) & BF, SF \\
90101005002 & 57193.6 & 38.8 & 0.063(32) & 0.8(5) &  &  &  &  & BF, SF \\
90201037002 & 57641.5 & 82.5 & 0.265(35) & 440(130) &  &  &  &  & P \\
90202038002 & 57668.8 & 46.0 &  &  & 0.23(4) & 30(100) & 960(320) & 0.07(23) & BF, Q \\
90202038004 & 57722.6 & 45.0 &  &  & 0.213(13) & 222(10) & 166(31) & 2.7(6) & BF, Q \\
90901332002 & 60263.9 & 71.9 & 0.117(10) & 4.9(10) &  &  &  &  &  \\
90901333002 & 60275.3 & 53.7 & 0.139(8) & 6.2(9) & 0.157(10) & 131.0(19) & 33(5) & 7.7(13) & BF, Q \\
\enddata
\tablecomments{The Notes column indicates when observations are affected by background flaring (BF), when the flux is highly variable during the observation (HV), and when QPOs (Q) and/or pulsations (P) are detected by \citet{Bachetti2022} and Bachetti et al. in prep.}
\end{deluxetable*}

Table~\ref{tab:rnqpo}  presents the complete set of best-fit parameters derived from our modeling of the PDS described above.
Figure~\ref{fig:rms} shows the variation of the total absolute rms of the features with frequency.
The points with upper limits refer to models where the AIC criterion suggested an advantage in adding a model component, but the bootstrap procedure returned a 3-$\sigma$ confidence interval including zero.
We use total rms and not the more customary fractional rms because the total X-ray flux of M82 is the combination of many X-ray sources, and in particular of both \Mone and \Mtwo.
Therefore, the fractional rms of the features will change randomly based on which source is more luminous during each observation, while the absolute rms retains a physical meaning in terms of total variable luminosity, in counts/s.

From this visualization, two distinct features are clearly identified. One at frequencies below $\sim0.02$ Hz whose rms does not depend on frequency, and one above, with a clear correlation between rms and frequency, that can confidently be identified with the QPO from \Mone studied by \citet{Strohmayer2003, Dewangan2006,Mucciarelli2006} (see below for an additional test of the association). The increase of absolute rms with frequency follows an approximate $\nu^{1/2}$ law, which might imply that the frequency is increasing with increasing flux of \Mone (i.e. implying an approximately constant \textit{fractional} amplitude).
There seems to be no evident correlation of any of the features with the appearance of pulsations from \Mtwo\footnote{The pulsation information is mostly extracted from \citet{Bachetti2022}, plus an additional detection that will be published by Bachetti et al. in prep.}, which is probably due to the fact that the red noise is influenced by both sources.

\subsection{Energy dependence}\label{sec:qpoenergy}

\begin{figure}[htbp]
    \centering
    \includegraphics[width=\linewidth]{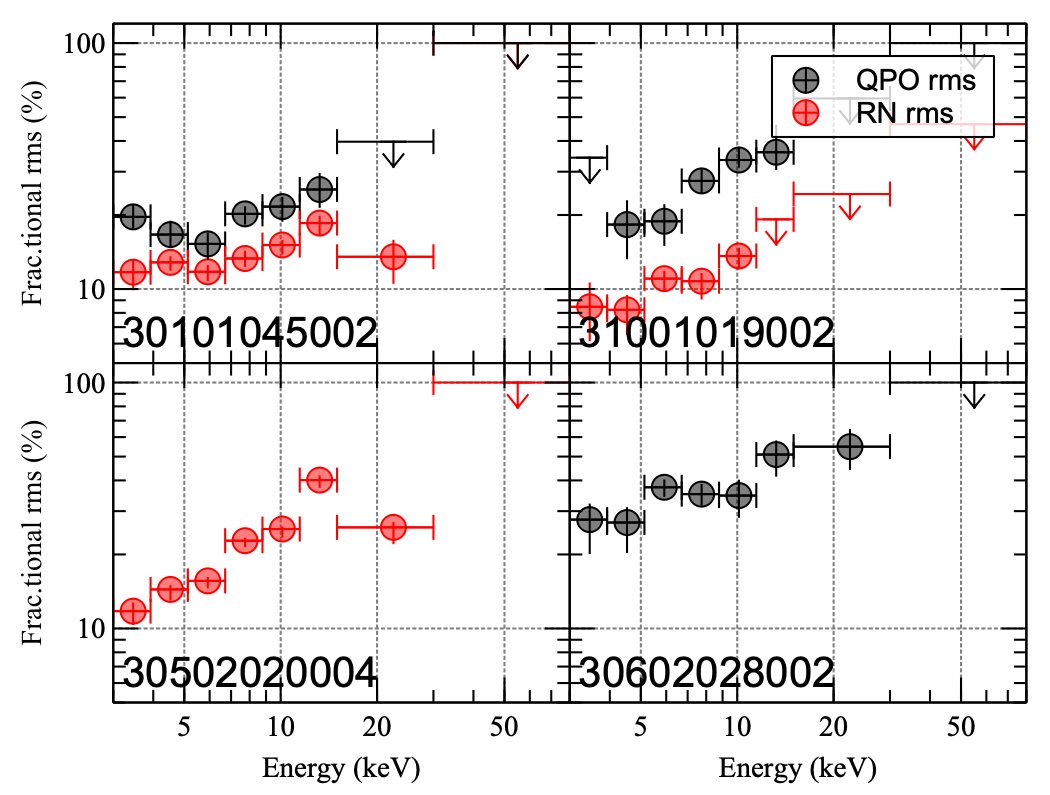}
    \caption{Fractional rms of the red noise and the QPO in different obsids. Horizontal bars with caps represent exact energy ranges, vertical bars are 1-$\sigma$ uncertainties. All variability is more significant at higher energies}
    \label{fig:rmsener}
\end{figure}
It is interesting to investigate how the features evolve with energy.
We divide the 3-80\,keV energy band into 8 intervals whose width follows approximately a geometric sequence but with larger intervals at higher energies to account for the very small number of high-energy counts.
We re-fit the model in each energy band; however, in most observations, the fit is not robust enough to leave all parameters free. Since there is no evidence of changes in the shape of the Lorentzian components with energy, only their normalization, and the Poisson noise level does not depart significantly from 2, we fix all parameters but the amplitudes of the one/two Lorentzians to their best-fit values from the total-flux analysis.
Through repeating the bootstrap procedure, we get sensible values for the amplitudes and their uncertainties, and we can calculate the rms at different energies.
Since we are proceeding on an observation-by-observation basis, we calculate the \textit{fractional} rms, as we are only interested in the relative variation of rms with energy.
The results are shown in Figure~\ref{fig:rmsener}, and clearly show that the rms increases with energy for all components.

\subsection{Search for other QPOs and harmonics}\label{sec:shiftandadd}
\begin{figure*}
    \centering
    \includegraphics[width=\linewidth]{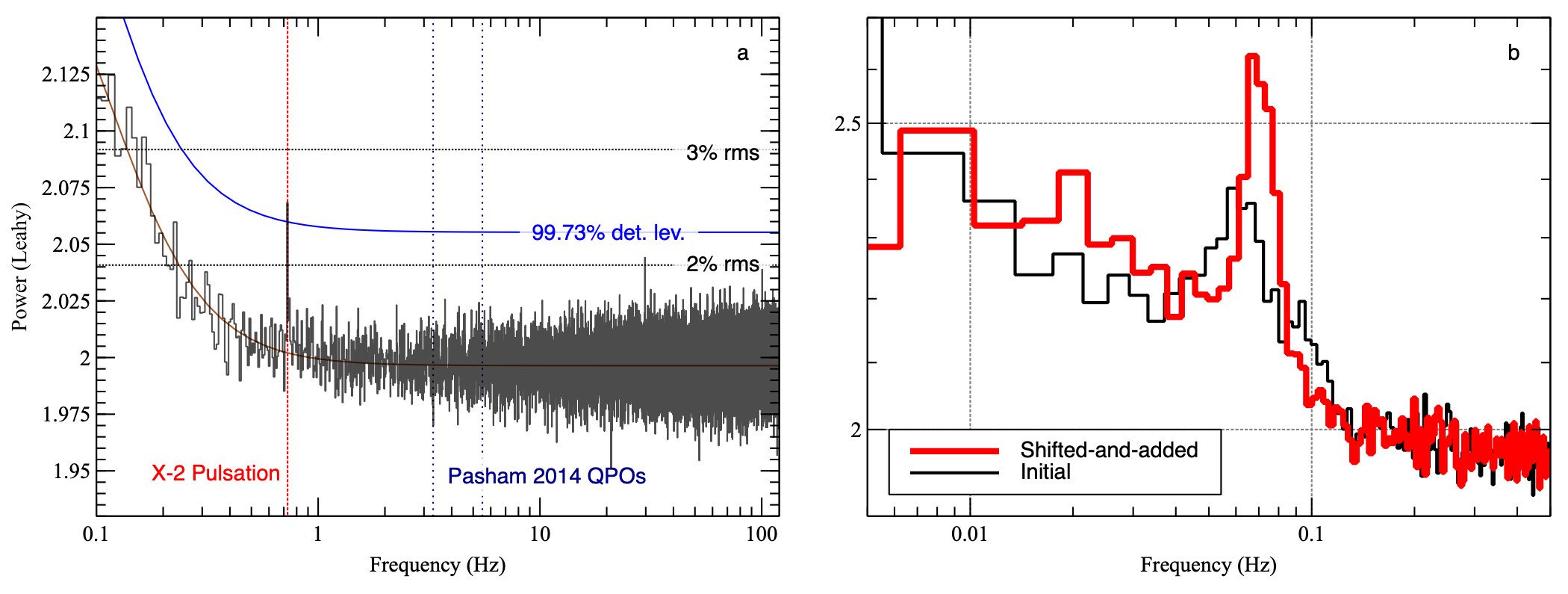}
    \caption{(a) Periodogram of the full $\sim5$\,Ms of \nustar observations of M82, including data intervals with poor star tracker coverage. The peak at $\sim0.7$\,Hz is the pulsation of \Mtwo, while the peak at $\sim30$\,Hz is not significant and sets the upper limit to any quasi-coherent variability to $\sim2\%$.
    (b) result of the shift-and-add technique applied to the 50-mHz QPO. The resulting Q factor is $\sim8$, and there is no significant harmonic at twice the frequency.}
    \label{fig:qposearch}
\end{figure*}
\citet{Pasham2014} reported the detection of 3- and 5-Hz QPOs from M82, most probably from \Mone, using \rxte data.
We now know that there are $\sim$15 ULXs between the galaxies M82, M81 and the satellite Holmberg IX \citep{waltonMultimissionCatalogueUltraluminous2022}, all within the $1^o$ field of view of the \rxte/PCA instrument used to detect the QPOs. Despite \Mone being the brightest at its maximum, all bright ULXs in the field, including \Mone, \Mtwo, Holmberg IX X-1, are known to be transient.
This raises the possibility that \Mone is not the source of the twin QPOs, and more generally, implies that the 3--5\% fractional-rms upper-limit derived by \citet{Pasham2014} from the blended light curve almost certainly underestimates the intrinsic rms amplitude that \Mone\ itself would need to produce the twin QPOs.

Despite the very long total exposure by \nustar, reproducing the result on \nustar data might not be straightforward, for various reasons. Despite the focusing capabilities that reduce considerably the contamination compared to \rxte, the \nustar PSF still includes \Mone, \Mtwo and at least one other ULX, M82 X-3. 
Also, \nustar has a much lower effective area than \rxte.
With these caveats in mind, we set out to look for the twin QPOs by averaging all M82 observations analyzed above.
This time, we were less worried about low-frequency leakage given the relatively clean part of the periodogram above 1 Hz, so we used all data, including those from intervals with poor star tracker coverage (see section~\ref{sec:datared}). 
Given the circumpolar position of M82, this allowed to almost double the exposure, improving the detection sensitivity by a factor $\sim\sqrt{2}$.
As can be seen in Figure~\ref{fig:qposearch}, there is no significant feature at frequencies above the pulsation frequency of \Mtwo with an rms reaching as low as $2\%$.

Additionally, we looked for harmonics of the 50-mHz QPO, in order to compare it to other known classes of QPOs such as Type-C from BH LMXBs. 
We used the shift-and-add technique \citep{barretHighCoherenceKHz2005} on the longest observing span available, the series of ObsIDs 800020920{02..11} from 2014. 
We split the observation in 128\,s intervals, calculated the periodogram in each, then averaged 25 such intervals to gain in signal to noise, creating a series of periodograms, each from 3.2\,ks of non-contiguous data. 
These were sufficient to track the evolution of the QPO frequency $\nu_{\rm QPO}$  during the observations.
We calculated the average QPO frequency $\bar\nu_{\rm QPO}$ and then shifted each periodogram by an amount corresponding to the $\nu_{\rm QPO} - \bar\nu_{\rm QPO}$ , obtaining the average shape of the QPO.
This technique was used successfully to characterize kHz QPOs in NS LMXBs, and in some cases even to discover the upper kHz QPO \citep{barretHighCoherenceKHz2005}. 
We also used a variation of this technique: since we were not looking for a frequency following a parallel track, but for a harmonic, we shifted the periodogram around $2\bar\nu_{\rm QPO}$, by a double amount.
We did not find evidence for a harmonic of the QPO.

\section{Discussion} \label{sec:discussion}
\begin{table}[htb]
\centering
\begin{tabular}{ccccc}
\hline
\hline
ObsID & Instrument & Date & Exposure &  Simultaneous \\
 &  &  MJD  & ks & NuSTAR ID\\
\hline
16580 & ACIS-S & 56691.8 & 47.5 & 80002092007 \\
17578 & ACIS-S & 57038.6 & 10.1 & 50002019002 \\
16023 & ACIS-S & 57042.0 & 10.1 & 50002019004 \\
18064 & ACIS-I & 57483.7 & 25.1 & 80202020006 \\
18068 & ACIS-I & 57502.8 & 25.1 & 80202020008 \\
18070 & ACIS-I & 57669.0 & 25.1 & 90202038002 \\
18072 & ACIS-I & 57723.4 & 25.6 & 90202038004 \\
26664 & ACIS-S & 60274.5 & 40.1 & 90901333002 \\
\hline
\end{tabular}
\caption{Quasi-simultaneous \chandra and \nustar observations used for the QPO identification.}
\label{tab:xmatch}
\end{table}

Our analysis is based on a multi-Lorentzian fit of the periodogram, as is often done in similar studies from accreting sources \citep{belloniAtlasAperiodicVariability1990}.
One difference is that we cannot rely on fractional variability when we study the evolution of the QPO, as we cannot have a clean view on \Mone using \nustar due to the presence of \Mtwo, and the source flux is unknown.
Hereafter, we will use fractional rms only to compare the strength of the QPO at different energy bands in a given observation, while we will use the absolute rms (in counts per second) when discussing the evolution over time.

This source confusion, on a related note, also hinders us from doing detailed modeling of the spectral break and its relation with the QPO frequency (\`a la \citealt{atapinUltraluminousXraySources2019}), because the red noise component is produced by both sources.

\begin{figure*}
    \centering
\includegraphics[width=0.48\linewidth]{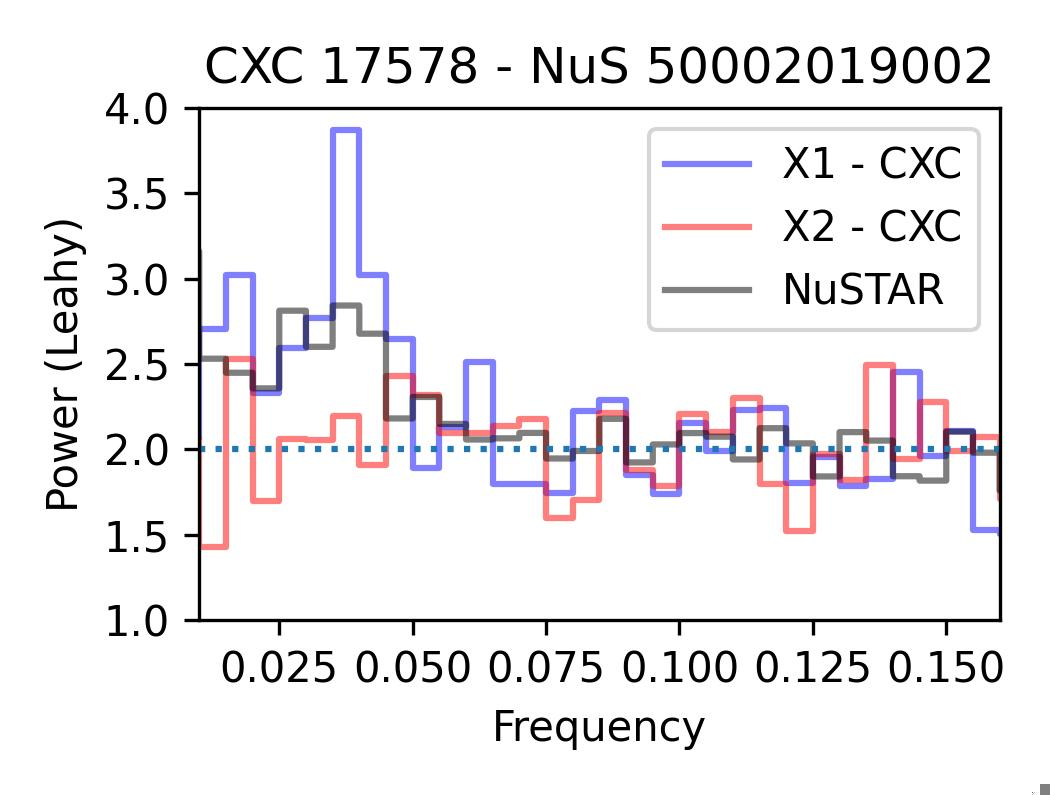}
\includegraphics[width=0.48\linewidth]{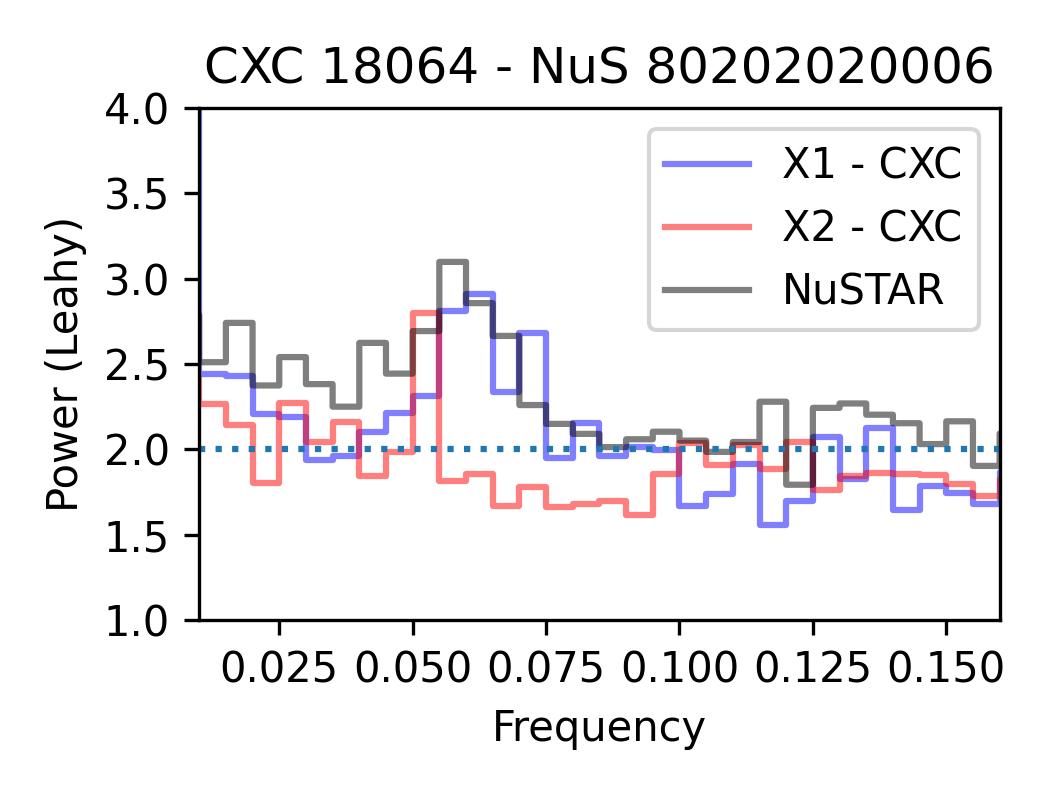}
    \caption{Simultaneous detections of the QPO in \nustar and \chandra data, showing a clear association with \Mone}
    \label{fig:qpox1}
\end{figure*}

\begin{figure}
    \centering
    \includegraphics[width=\linewidth]{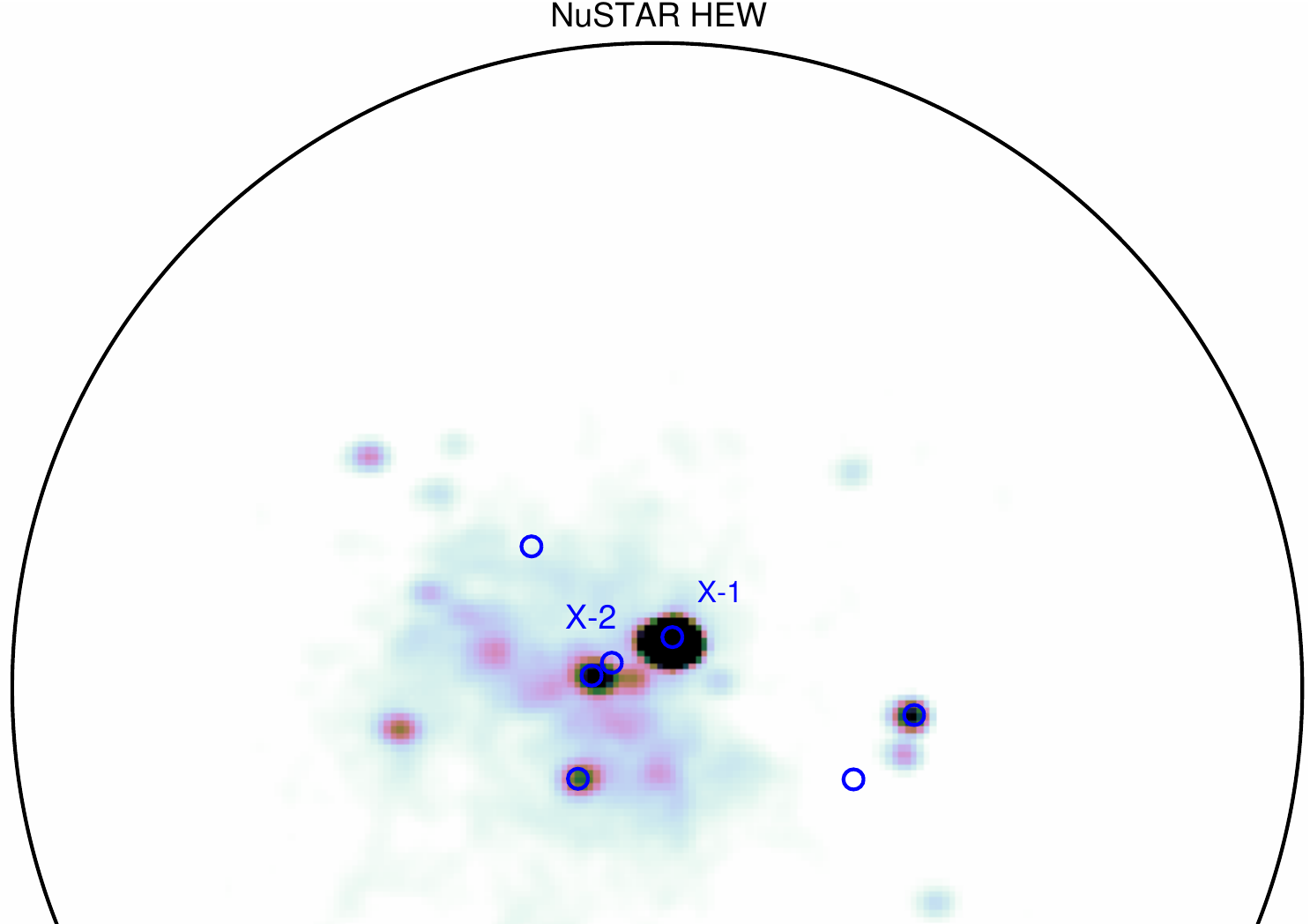}
    \caption{Image of \chandra ObsID 18072, simultaneous to \nustar ObsID 90202038004. Blue circles indicate the 7 ULXs in the catalogue by \citet{liuUltraluminousXRaySources2005}. The QPO is significantly detected in \nustar data and not in \chandra data of any source, but the \chandra image shows that the emission is dominated by \Mone, while \Mtwo is undetected.}
    \label{fig:18072}
\end{figure}

The identification of the QPO with \Mone is tricky by itself. Most detections come from missions that do not resolve the two ULXs \citep{Strohmayer2003}, such as \xmm and \rxte, although with a good degree of confidence given the higher flux of \Mone and the change of the QPO strength when carefully selecting data closer to \Mone in \xmm \citep{feng2007origin}.
We looked for observations having quasi-simultaneous \chandra observations showing a low state from \Mtwo and the QPO in the data, and searched the \chandra data themselves for QPO detections.
Most on-axis \chandra observations of M82 are often plagued by pileup and the sensitivity to any variability was low.
We found that the QPO was present in \nustar ObsID 90202038004, with \chandra ObsID 18072 showing a low state of \Mtwo, and, for the first time, a tentative detection in off-axis \chandra ObsIDs 17578 and 18064, this time using events firmly associated with \Mone, at a frequency compatible with the detection from the simultaneous \nustar ObsIDs (Figs.~\ref{fig:qpox1} and~\ref{fig:18072}).

The evolution of the $20-300$\,mHz QPO from \Mone observed in our 10-year \nustar campaign shows behavior compatible with what was previously observed for this source \citep{Mucciarelli2006,feng2007origin,atapinUltraluminousXraySources2019}, extending the range of observed frequencies.
The low background and hard response of \nustar, together with our power spectral cleaning procedure, allows for a better modeling of the red noise component of the power spectrum and a systematic analysis of the evolution of the QPO using almost 3 Ms of exposure.

Low-frequency QPOs are observed in all classes of accreting sources \citep{wijnandsBroadbandPowerSpectra1999a}.
In particular, NS and BH binaries have a number of low-frequency features spanning the frequency range $\sim$0.001--50\,Hz.
Some evolution of the QPO frequency over time is a hallmark of almost all classes of QPOs \citep{vanderklisRapidXrayVariability2006,motta2016review}, and in particular, all classes of low-frequency QPOs.
Typically, the absolute rms is linearly correlated with flux, with the fractional rms being stable or slightly decreasing.
Also the QPO frequency is often seen correlating with the flux on short time scales, while the correlation is broken on long time scales \citep[the so-called ``parallel tracks''][]{vanderklisRapidXrayVariability2006}.
At higher energies, the characteristic frequency does not change significantly, while their fractional rms generally increases.

One class of QPOs that can naturally be compared with ours is Type-C QPOs from BH LMXBs, the most common oscillatory pattern in this class of sources.
They start to appear in the low-hard state and their frequency generally increases as the luminosity (and probably the mass accretion rate) increases, going through the intermediate states that lead to the high-soft state and sometimes the so-called ultraluminous state. These intermediate states are also associated with the presence of transient jet ejections.
Type-C QPOs are generally also accompanied by broad red noise (or flat-top) components, usually also modeled with Lorentzians whose characteristic frequencies (Eq. \ref{eq:lorentz}) evolve in parallel with that of the QPO, and one or more harmonics.
Models for these QPOs often involve Lense-Thirring precession around a rotating BH, and put their frequency in relation with other oscillatory components such as broadband noise or high-frequency QPOs.
These models use different approaches from single-particle motion (the original relativistic precession model, or RPM; \citealt{stellaKHzQuasiperiodicOscillations1999}), to precessing rings \citep[e.g.][]{psaltisOriginQuasiPeriodicOscillations2000}, to entire regions of the disk that precess like a solid body (\citealt{Fragile2007,ingram2011model}, see also
\citealt{ingram2019review} for a review).

Interpreting \Mone's QPO as a Type-C QPO, one can also be tempted to go one step further: all time scales around a gravitating body scale with mass, and notably this includes orbital frequencies. The fact that this QPO are about an order of magnitude slower than typical Type-C QPOs leads to a mass estimate of an order of magnitude above stellar-mass BHs, in the regime of small intermediate-mass BHs.
This kind of scaling is often attempted, using different variability components, and in the case of \Mone this has often led to claims of IMBH origin for this source \citep[e.g.][]{Strohmayer2003, Pasham2014}.

However, similar claims were made for other sources. One such source is \Mtwo, for which \citet{feng2010discovery} estimated a mass of 12,000–43,000 \msun by rescaling the \chandra-detected mHz QPOs to Type-C QPOs. 
However, this famously turned out to be a pulsar.
It is generally difficult to compare QPO phenomena from different objects, and there is a wealth of QPO phenomena on many time scales in stellar mass compact objects, including at even lower frequencies.
One example is the mHz QPO observed in low-mass NS and BH X-ray binaries \citep[e.g.][]{revnivtsevNewClassLow2001,xiaoSystematicStudyMillihertz2024}.

Moreover, we note that the spectrum of \Mone does not depart significantly from the bulk of ULXs \citep{brightmanSpectralEvolutionUltraluminous2020}, that are increasingly identified as super-Eddington accreting stellar-mass objects. 
In some cases the detection of pulsations that univoquely identify them as NSs, but in general, their spectral shapes do not resemble the bulk of sub-Eddington accreting objects (\citealt{Gladstone2009MNRAS.397.1836G}), and frequently show signatures of the strong winds expected from super-Eddington accretion \citep{Middleton2014,Middleton2015b,Pinto2016Natur.533...64P,Pinto2017,Kosec2021,pintoUltraluminousXraySources2023}.
In particular, NGC 5907 X-1 has a comparable flux to \Mone despite being a NS \citep{israelAccretingPulsarExtreme2017,furstProbingNatureLow2023}.

In addition, the decrease of the coherence of the \Mone QPO with frequency observed here (Fig. \ref{fig:rms}) is puzzling, and unlike what is observed in Type-C QPOs where the coherence tends to generally increase with frequency, at least up to the onset of the soft state where these oscillations disappear. Admittedly, this might be an observational bias: the integration time required to detect the QPO might just be longer than the variation time for the frequency. We need thousands of oscillations of the QPO before being able to detect it, and if the frequency is changing rapidly this would artificially increase the measured width, and so its quality factor.
A clear example of this are ObsIDs 80002092006 and 80002092007, where using the shift-and-add technique clearly improves the quality factor (Section~\ref{sec:shiftandadd}). 
In an effort to determine if this played a role, we used the shift-and-add technique in all observations with a strong QPO and at least $\sim$90\,ks exposure. The technique always improved the $Q$ factor throughout the frequency range, but without significant advantages for the higher frequencies. 

Interestingly, bright accreting pulsars also show low-frequency QPOs, with a phenomenology and characteristic frequencies similar to what we observe here.
For these sources, low-frequency QPOs are often hypothesized to arise from the Keplerian frequency at some important radius, or from a beat between the Keplerian frequency at the truncation radius and the spin frequency (beat frequency model, \citealt{alparGX5MillisecondPulsar1985}). \citet{manikantanEnergyDependenceQuasiperiodic2024} provides a table of the energy-dependent QPO parameters for a number of sources over multiple observations. V0332+53 is the only one with sufficient observations to see the evolution of the quality factor over a wide range of QPO frequencies, and $Q$ is interestingly seen to decrease between 10 and 2 with increasing frequency, as we observe for \Mone. 
As a general rule, we argue against using QPOs alone to infer the mass of accreting objects (cf \citealt{Middleton2011MNRAS.411..644M}). The QPOs studied in this work are compatible with phenomenology observed in accreting sources of different kinds, including NSs.

The non-detection of the twin 3--5\,Hz QPOs reported by \citet{Pasham2014} can be explained in various ways. First of all, the oscillations might be transient, and have disappeared over time: the result by Pasham integrated many years of data.
It is also possible that the filtering of ``flaring'' observations by \citet{Pasham2014}, which reduced considerably the time intervals analyzed to obtain their average periodogram, was more aggressive than ours, and that we need more observations without flaring or high variability to detect those QPOs with \nustar.
Another possibility might even be that the source of QPOs is \textit{not} in M82 after all: the field of view of \rxte is one degree, which includes three ULX host galaxies (M82, M81, Holmberg IX) with at least 9 ULXs \citep{liuUltraluminousXRaySources2005}. 

\section{Conclusions}
We made an extensive, 10-year study of the QPOs from \Mone using \nustar.
Using archival \chandra data, we made a robust identification of the 50-300 Hz QPO with \Mone, which confirms previous evidence.
Thanks to the sensitive response at and above 10 keV, we were able to detect the QPO over $\sim$70\% of the 3 Ms of existing M82 observations, characterizing its behavior.
We used a novel approach to cleaning the periodogram in order to better fit the low-frequency component and get reliable fit parameters for the QPO as well.
The QPO tends to decrease its coherence as its frequency increases, but it is not clear whether this is due to fast variations of the frequency that we are not able to follow due to the long exposure required by the detection at these count rates.
We note that LFQPOs in this range of frequencies and with similar behavior are observed in many accreting systems, including NSs, so that any inference on the mass of the compact object based on the frequency of the QPOs should be taken with a grain of salt.

\software{astropy \citep{2013A&A...558A..33A,2018AJ....156..123A,astropycollaborationAstropyProjectSustaining2022},
          Stingray \citep{huppenkothenStingrayModernPython2019},
          CIAO \citep{fruscioneCIAOChandrasData2006},
          Astroquery \citep{ginsburgAstroqueryAstronomicalWebquerying2019}
          }
\begin{acknowledgements}
The authors wish to thank Phil Uttley for discussions on the spectral cleaning procedure.
MB was supported in part by the Italian Research Center on High Performance Computing Big Data and Quantum Computing (ICSC), project funded by European Union - NextGenerationEU - and National Recovery and Resilience Plan (NRRP) - Mission 4 Component 2 within the activities of Spoke 3 (Astrophysics and Cosmos Observations).
DJW acknowledges support from the Science and Technology Facilities Council (STFC; grant code ST/Y001060/1).
\end{acknowledgements}

\appendix

\section{Periodogram production and filtering}\label{sec:periodogram}
\begin{figure*}
    \centering
    \includegraphics[width=0.85\linewidth]{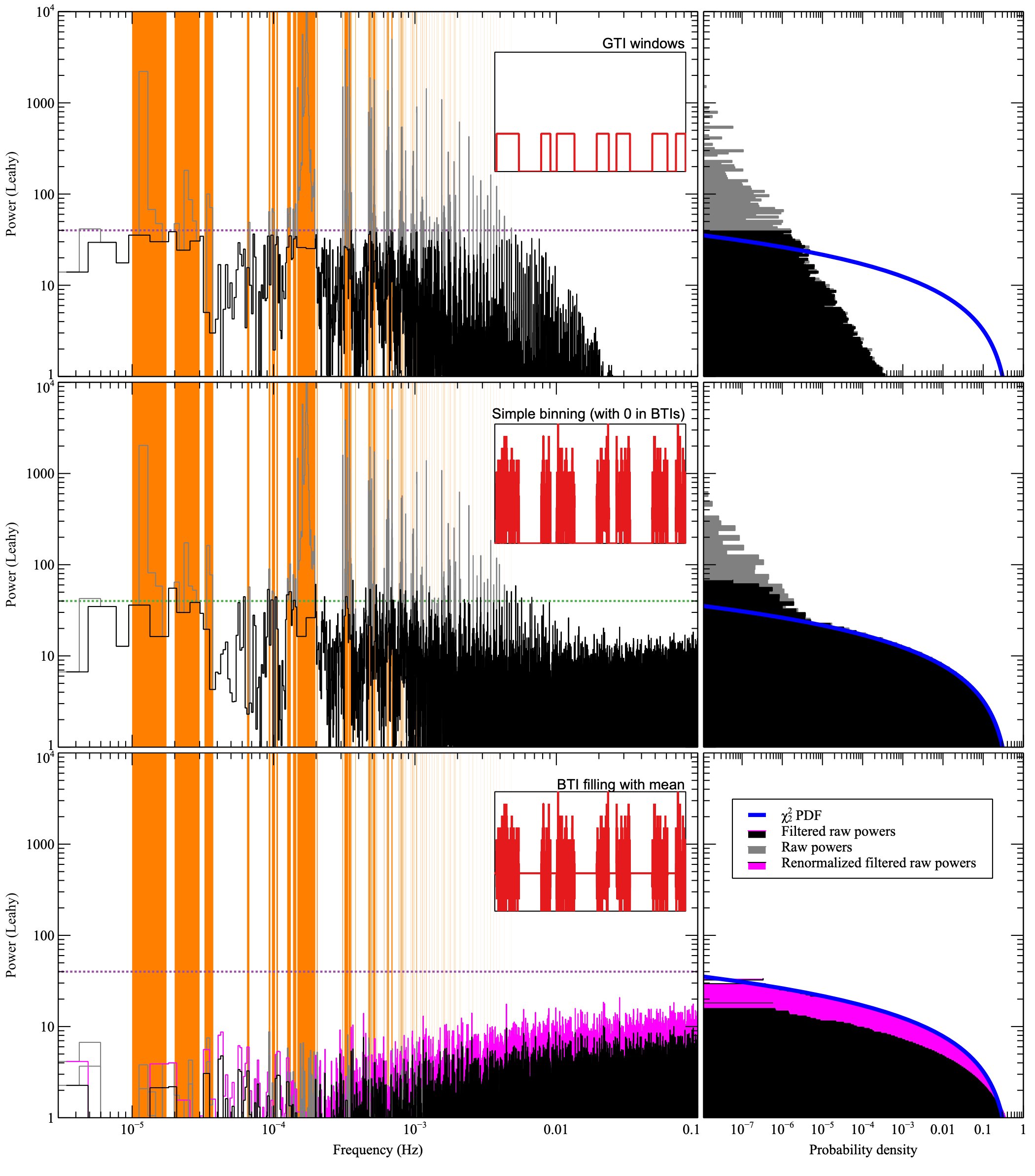}
    \caption{Procedure to filter periodograms from the effect of visibility windows described in Section~\ref{sec:periodogram}, using simulated Poissonian data (so, no source variability) with the same mean count rate and GTIs of ObsID 80002092006.
    (Top) Periodogram of the visibility light curve, showing the features corresponding to the missing data. We set a threshold and select a number of bad frequency intervals (orange) to be black listed.  (Middle) Periodogram of the binned light curve with 0 outside GTIs: it contains many of the the same features, with similar powers, plus the expected white noise from the data. After notch filtering, some powers clearly exceed the expected $\chitwo$ distribution. (Bottom) Periodogram of the binned light curve with the mean counts per bin used as a filler outside GTIs instead of 0. Most of the features disappear from the periodogram even before notch filtering, and the powers follow the correct distribution, but with the wrong normalization, which is corrected as described in the text}
    \label{fig:pds_clean}
\end{figure*}

Our source light curve is produced by sampling the photons from the source region falling into equispaced time intervals with resolution $t_{\rm samp}$.
Alongside the source light curve, we create synthetic visibility light curves with the same sampling interval that are equal to the mean counts per bin of the source light curve during GTIs and 0 outside.
We take a periodogram of this synthetic light curve, that shows strong features corresponding to the missing data while going rapidly to zero at high frequencies, because it does not contain noise.
In this periodogram, we can set a threshold, and single out the strongest peaks, to be used for notch filtering.
The idea is that we will eliminate these frequencies from the final periodogram, and then make a geometrical rebinning (as it is commonly done) that will average the remaining nearby bins, maintaining the statistical properties of the periodogram.

However, if we take the source light curve at face value, using 0 outside GTIs, we can see that if one applies the notch filter above straight away, a large number of powers do not follow the expected distribution (Figure~\ref{fig:pds_clean}, middle panels).
This is because some noise is still leaking at frequencies near the blacklisted ones.

But we can also fill the bad time intervals of the light curve not with zeros, but with the mean of the data. This produces a much smaller effect of the visibility windows on the final periodogram, creating an almost flat distribution of power, and the notch filter will just be an additional cautionary measure (Figure~\ref{fig:pds_clean}, bottom panels, black data).
In a real-life situation, when there is some long-term source variability, filling bad intervals with the mean will not be as clean as this example, so the notch filtering will be useful.
There is an additional measure to take, however.
The periodogram is a measure of the variance of the data, and it is calculated from an FFT which contains a division by total number of data points. However, the filled data points do not contribute to the variance, and this means that the calculated power will be lower than the expected value. To reinstate the correct normalization of the white noise, we need to multiply this periodogram by $n_{\rm tot}/n_{\rm gti}$, where $n_{\rm tot}$ is the total number of bins and $n_{\rm gti}$ the number of bins in GTIs
(Figure~\ref{fig:pds_clean}, bottom panels, magenta data).
Additionally, the measured rms of signal will still be underestimated by another factor $n_{\rm tot}/n_{\rm gti}$.

To verify that the method does not alter the response of the periodogram in a frequency-dependent way, we performed the following test: we generated synthetic light curves with a sinusoidal modulation at 10\% fractional amplitude, using the same GTIs and mean flux as ObsID 80002092006. The light curve had no Poisson noise, only the smooth sinusoidal modulation.
Outside GTIs, we filled the time intervals with the mean value of the flux as we did for the data.
We repeated the experiment for 1000 frequencies distributed log-uniformly over the range $10^{-4}-1$ Hz.
We always used frequencies extracted from the grid of the periodogram, in order to avoid the expected $\mathrm{sinc^2}$ response degradation when moving away from the center of the frequency bin \citep{van1989fourier}.
We applied the same notch filters as the real data, and we measured the decrease of rms (hereafter, the damping factor) in the remaining frequencies compared to an uninterrupted pulsation over the whole observation.
The decrease of rms was the same at all frequencies and exactly what expected from the fact of having missing data (i.e., outside GTIs there is no variability and the total variance decreases).
Only very close to the notch-filtered frequencies, we noticed some ``wiggles'' in the damping factor, by 10\% at most.

\section{Alternative periodogram analysis: Bartlett + Lomb-Scargle}
Another widely used tool for detecting and characterizing periodic signals is the Lomb-Scargle periodogram \citep{lombLeastsquaresFrequencyAnalysis1976a,scargleStudiesAstronomicalTime1982a}, which has proven to be more effective in detecting periodic patterns even when observations are unevenly spaced, providing a reliable solution for analyzing time series data in astronomy and diverse scientific fields as well. The Lomb-Scargle allows for covering a wide range of low frequencies in comparison to the Bartlett periodogram, allowing for a more comprehensive analysis of low-frequency signals.

The only downside of this periodogram is that its powers are not guaranteed to be uncorrelated, and the assumption of $\chitwo$-distributed powers is not as solid as for the periodogram.
For a comprehensive view on modern algorithms to compute the Lomb-Scargle periodogram and their limitations, see \citealt{vanderplas2018understanding}.

Nonetheless, the Lomb-Scargle periodogram is very useful for the analysis of unevenly sampled data, including light curves with missing data.
Our original approach to the analysis of the datasets in this paper was a hybrid approach, using the Bartlett periodogram and the Lomb-Scargle periodogram at the same time.
We write it here because it could be an interesting inspiration for future works and, in any case, it represents an alternative (even if not completely independent) approach to the analysis presented in this paper.
We calculated the Lomb-Scargle periodogram simply discarding light curve bins outside GTIs, and avoiding oversampling (i.e., using the same spectral resolution of the FFT), which limited the correlations between powers. Our sample time was 0.1, giving 5 Hz as the Nyquist frequency, but we found that the periodogram departed significantly from the expected noise level of 2 when reaching about half the Nyquist frequency. Otherwises, the Lomb-Scargle and the Bartlett periodograms have very good overlap in the common frequency ranges, and we decided to eliminate the frequencies above 1 Hz from the Lomb Scargle periodogram for extra caution. We checked that the assumption of $\chitwo$ distributed powers is justified for our case by plotting the distribution of the Lomb-Scargle powers of simulated and real data similarly to~\ref{fig:pds_clean}.
Moreover, even though in theory it should not have been needed, we blacklisted the frequencies corresponding to the orbital occultations similarly to the description in Appendix~\ref{sec:periodogram}.
The raw periodograms have very different frequency resolutions, but we rebinned both periodograms with geometrically increasing frequency bin sizes, and defined a threshold frequency $\nu_{\rm thr}$ where the frequency resolution of the rebinned Lomb-Scargle periodogram reached the frequency resolution of the Bartlett periodogram.
From that point on, we used a hybrid periodogram containing the Lomb-Scargle powers below $\nu_{\rm thr}$ and the Bartlett ones above. The periodogram was characterized by the power values and the number of averaged powers, either from rebinning or -- in the case of the Bartlett one -- averaging of multiple periodograms.

We proceeded to fit the multi-Lorentzian model of Section~\ref{sec:bootstrap}, with the same methods, to this hybrid periodogram. Indeed, the results using this alternative method are compatible with those obtained for the single filtered periodogram in Section~\ref{sec:bootstrap}.

\bibliographystyle{aasjournal}

\end{document}